\documentclass[a4paper]{amsart}
\usepackage{amssymb,amsthm,euscript,amsmath,graphicx,array}

\newtheorem{thm}{Theorem}[section]
\newtheorem{lem}[thm]{Lemma}

\newtheorem{rem}{Remark}[section]

\newtheorem{exam}{Example}

\begin{document}
\pagestyle{plain}

\title{Weakly nonlocal continuum physics -
    the Ginzburg-Landau equation}
\author{Peter V\'an\\ }
\address{Budapest University of Technology and Economics\\
Department of Chemical Physics\\
1521 Budapest, Budafoki \'ut 8.}

\email{vpet@phyndi.fke.bme.hu}

\date{\today}

\begin{abstract}
Some peculiarities of the exploitation of the entropy inequality in
case of weakly nonlocal continuum theories are investigated and
refined. As an example it is shown that the proper application of the
Liu procedure leads to the Ginzburg-Landau equation in case of a weakly
nonlocal extension of the constitutive space of the simplest internal
variable theories.
\end{abstract}

\maketitle

\section{Introduction}

Ginzburg-Landau equation and its variants appear in different fields of
physics and are applied to several phenomena. Its physical content and
the way to obtain it is sound and transparent. The traditional
derivation of the equation comprises two main ingredients (see e.g.
\cite{PenFif90a})
\begin{itemize}
\item The static, equilibrium part of the equation is derived from a
variational principle.
\item The dynamic part is added by stability arguments (relaxational
form).
\end{itemize}

The two parts are connected loosely and in an ad-hoc manner. As a
classical field equation defined on nonrelativistic space-time, the
Ginzburg-Landau equation should be compatible with the general balance
and constitutive structure of continuum physics. Recently there has
been several efforts to give a uniform reasoning of the equation on
pure thermodynamic ground and to generalize the method of the
derivation \cite{Gur96a,Mar02a,Van02a1,Van03a1}. The treatment of
Ginzburg-Landau equation is a kind of test of weakly nonlocal
(gradient) thermodynamic theories.

In some previous works the possible role of internal, dynamic variables
in weakly nonlocal extensions of the constitutive state space was
investigated both with the heuristic method of classical irreversible
thermodynamics and with the more exact Liu procedure
\cite{Van02a1,Van03a1}. It was found that with the help of special
dynamic variables (current multipliers) appearing in a generalized
entropy current one can find an equation that is similar to the
Ginzburg-Landau equation, but is not the same. The equation was called
{\em thermodynamic Ginzburg-Landau} equation. The static part of the
Ginzburg-Landau equation gives nonhomogeneous solutions in case of
uniform boundary conditions, therefore it is a pattern forming
equation. On the other hand, thermodynamic Ginzburg-Landau is not a
pattern forming equation, its corresponding solutions are homogeneous.
This is an important difference and corresponds well to the fact that
with internal variables one introduces a local theory, all the
functions in the basic state space are local. Therefore any
nonlocalities formulated by an internal variable theory are {\em
relocalized} in this sense. Pattern forming theories, like those based
on the Ginzburg-Landau equation, cannot be relocalized.

In this paper it is shown, that there is a non-equilibrium
thermodynamic approach also for pattern forming equations, however, it
is not relocalizable, independent of any kind of internal variables. A
pure thermodynamic derivation can unify the mentioned two parts of the
derivation, there is no need to postulate any kind of variational
principle. The Euler-Lagrange form of the static part of the equation
turns out to be a consequence of the Second Law.

In the paper the key ingredients of the methodology and the
mathematical background to exploit the Second Law inequality is shortly
summarized in weakly nonlocal continuum theories (gradient theories,
theories with coarse grained thermodynamic potentials, phase-field
models, etc..). After the introduction, an overview of the methods to
exploit the Second Law inequality is given. We call the attention to
the essential differences between the local and weakly nonlocal
theories and a constructive way to arrive solvable Liu equations is
given. The most important difference is that the weakly nonlocal
extension of the constitutive state space, implies that some space
derivatives of the constraints (evolution equations) are to be
considered as constraints, too. A simple example show the application
of Liu procedure. In the following section Liu procedure is applied to
derive the Ginzburg-Landau equation. Different kind of generalizations
are treated, the difference between the relocalized and pattern forming
(non relocalizabile) theories is formulated. A new, generalized
formulation of Liu's theorem and its direct proof  from Farkas' lemma
is given in an Appendix.

\section{Methods to exploit the Second Law inequality - procedures of CIT,
Coleman-Noll and Liu}

In every non-equilibrium thermodynamic theories an important
theoretical problem is to formulate the correct form of the evolution
equations taking into account the requirement of the entropy
inequality. The most predictive methodological solution of the problem
is far from being trivial and originated from Coleman and Mizel. They
essentially reverse the way of thinking: one should look for the
solution of the entropy inequality taking into account the evolution
equations as constraints \cite{ColMiz64a,MusEhr96a}. In continuum
physics the dynamic equations are given in a determined form (e.g. as
balances of extensives), except some constitutive, material functions.
The task is to ensure the nonnegativity of the entropy production with
appropriate constitutive assumptions. Therefore one should specify the
undetermined material functions such, that in case of all possible
solutions of the dynamic equations the form of the constitutive
functions, the material properties, ensure the nonnegativity of the
entropy production. In case of weakly nonlocal theories the entropy
current plays a distinguished role. The entropy and the entropy current
are both constitutive and are to be determined according to the above
requirement (as it was suggested in extended rational thermodynamics
\cite{MulRug98b}). The entropy possibly should preserve its potential
character in a general sense, therefore solving the above problem (e.g.
Liu equations) the practical aim is a simplification in such a way that
every constitutive quantity including the entropy current could be
calculated from the entropy function.

There are three basic methods to exploit the entropy inequality.
\begin{itemize}
\item \textit{Heuristic.} This is the force-current method of classical
irreversible thermodynamics (CIT). In classical problems, in case of
simple state spaces, the form of the entropy production is quadratic
and the inequality can be solved. The method can be justified by the
Liu procedure in case of the traditional, simplest state spaces
\cite{Van03a1}, and the method can be generalized to include
non-classical entropy currents \cite{Ver83a,Nyi91a1}.

\item \textit{Coleman-Noll procedure.} In the Coleman-Noll procedure
one exploits the constraints (e.g. dynamic equations) directly,
substituting them into the entropy inequality. The degenerate form of
Liu's theorem (T\ref{degLiu}) is applied. One usually assumes a
specific form of the entropy current. There are essentially two choices
here. The entropy current can be the classical (${\bf j}_s={\bf
j}_q/T$), or a generalized one. In weakly nonlocal considerations both
classical (see e.g. \cite{Gur96a,CimKos97a}) and generalized forms are
applied. Generalizations of the entropy currents (or currents of other
thermodynamic potentials) can be suggested on different grounds and
they give good results with the procedure \cite{MarAug98a,Mar02a} .

\item \textit{Liu procedure.} With Liu procedure one applies Liu's
theorem  with Lagrange-Farkas multipliers (T\ref{Liu} in the Appendix).
At the first glance the application of this method seems to have only
practical advantages. However, as the Lagrange multiplier method
preserves the simple form of the constraints in question in extremum
problems, with Lagrange-Farkas multipliers one can preserve and exploit
the structure of the constraint and the entropy inequality. The
question is not purely mathematical, because there are cases where the
multipliers cannot be eliminated and they can get physical
significance. Moreover, an inevitable advantage of Liu's method is that
the structure of entropy inequality makes possible to solve completely
the physical problem.

The train of thought is the following. The entropy current is
considered as an independent constitutive quantity. With a proper
choice of the constitutive space we can solve the Liu equations and
determine the entropy current. Hence the entropy inequality simplifies
considerably. The point of view of Onsagerian CIT is important here:
with a proper identification of thermodynamic currents and forces the
resulted entropy inequality can be solved, determining all constitutive
quantities. This ensures, that our theory is independent of further
artificial constraints, the entropy inequality becomes a consequence of
material properties.
\end{itemize}

In weakly nonlocal continuum physical calculations with Liu procedure
one should consider some additional practical rules. There the
constraints are (partial) differential equations. The functions in the
differential equations form the {\em basic state space}. The
constitutive quantities depend on these functions, on the {\em basic
state} and on some of its derivatives. These derivatives are locally
independent therefore the problem is algebraically manageable. The
basic state variables and some of its derivatives can be included into
the \textit{constitutive state space} (or simply state space
\cite{MusAta01a}), into the domain of the constitutive functions. The
entropy inequality with its special balance form determines the
independent variables of the algebraic problem: those are the
derivatives of the constitutive state, the so called {\em process
directions}. The choice of the constitutive state space is crucial and
determines the restricted constitutive functions, after applying Liu
procedure. the peculiarity of weakly nonlocal theories is that
depending on the particular state space, some space derivatives of the
original constraints (e.g. dynamic equations) further restrict the
process direction space, therefore they should be considered in Liu's
theorem as additional constraints. In the following we will show some
examples to clarify the most important practical rules in the
application of the formalism. One can find other examples on the
application of derivative constraints in \cite{HerAta98p,VanFul03m}.

\begin{rem} The mentioned exploitation methods of the Second
Law, being algebraic, are essentially independent of the solvability of
the dynamic equations, whether the associated problems are well or ill
posed, on the applied function spaces, etc... For example the number of
the equations can be less than the number of variables in the basic
state space. The {\em wanted fields}, the functions searched in the
final resulting differential equation can be different from the basic
state.
\end{rem}

\begin{exam}
In this example the {\em basic state space} is formed by two times
differentiable real functions $x: \mathbb{R}\mapsto \mathbb{R}$. The
{\em constitutive space} is spanned by the basic state and its
derivative $(x,x')$. We are looking for scalar valued differentiable
functions $F$ and $S$ as lying in the constitutive space so that
$$
    S'(x,x')\geq 0
$$
\noindent for all $(x,x')$ satisfying the {\em constraint}
\begin{equation}
    F(x,x')=0.
\label{cond1}\end{equation}

Evidently $S'(x,x') = \partial_1S x' + \partial_2S x''$, where
$\partial_n$ denotes the partial derivative according to the $n$-th
variable. Therefore, the space of the {\em process directions} (the
space of independent variables in Liu's theorem) is spanned by $x''$.
We are looking for conditions on $S$ and $F$ that the above inequality
should be true for all $(x,x')$ solving (\ref{cond1}), but
independently of the values of $x''$. The degenerate case of Liu's
theorem (T\ref{degLiu}) gives some conditions. The single Liu equation
is
$$
    \partial_2S = 0.
$$

Therefore $S$ is independent on $x'$. The dissipation inequality can be
written in the following simple form
\begin{equation}
\partial_1S x' = \frac{dS}{dx}(x)x' \geq 0
\label{entr1_ineq}\end{equation}

The above inequality does not give any condition for $F$. However, let
us observe, that one of our previous assumptions was too strong. The
process direction variable $x''$ is not really independent on the state
space, the derivative of (\ref{cond1}) gives a further restriction
\begin{equation}
\partial_1F x' + \partial_2F x'' = 0.
\label{cond2}\end{equation}

Considering this condition we apply Liu's theorem (T\ref{Liu}) with the
multiplier method, introducing the multipliers $\lambda_1$ and
$\lambda_2$ for the constraints (\ref{cond1}) and (\ref{cond2})
respectively
\begin{eqnarray*}
\partial_1S x' &+& \partial_2S x'' - \lambda_2(\partial_1F x' +
    \partial_2F x'') - \lambda_1 F = \\
&=& (\partial_1S - \lambda_2\partial_1F) x' + (\partial_2S -
\lambda_2\partial_2F) x'' - \lambda_1 F \geq 0
\end{eqnarray*}

Therefore we can read the Liu equation as follows
$$
\lambda_2 \partial_2F - \partial_2S = 0
$$

Expressing the multiplier and substituting into the dissipation
inequality we get
$$
\partial_1S x' - \lambda_2 \partial_1F x' - \lambda_1 F =
(\partial_1S - \partial_2S (\partial_2F)^{-1}\partial_1F) x' -
\lambda_1 F \geq 0
$$

In this example we face to a partially degenerate case, hence with
$\lambda_1=0$ we can give the general solution of the above inequality,
as
$$
\partial_1S - \partial_2S (\partial_2F)^{-1}\partial_1F =
L(x,x') x',
$$

\noindent where $L$ is nonnegative. Given a function $S$ we can
calculate $F$, with appropriate conditions on $L$. For example if
$S(x,x')= x\cdot x'$ and $L=constant$, then $F(x,x')=f(x^{L-1}x')$ is a
solution of the above equation for any $f: \mathbb{R}\mapsto
\mathbb{R}$.
\end{exam}

\section{Weakly nonlocal non-equilibrium thermodynamics --
Ginzburg-Landau and thermodynamic Ginzburg-Landau equations}

Let us denote an internal variable (e.g. an order parameter of a second
order phase transition) characterizing the microstructure of the
material by $\xi$. In this case the {\em Ginzburg-Landau equation} can
be written as
\begin{equation}
\partial_t\xi = -\gamma_1 \Gamma_\xi + \gamma_2 \Delta\xi,
\label{GL_eq}\end{equation}

\noindent where $\Gamma_\xi$ is the partial derivative of the
appropriate thermodynamic potential (e.g. $\Gamma_\xi = \partial_\xi
f$, where $f$ is the free energy), and $\partial_t$ denotes the partial
time derivative. Here we assumed that the internal variable $\xi$ is
not related to the mechanical motion, therefore the choice of the frame
(partial or substantial time derivatives) is irrelevant. $\gamma_1$ and
$\gamma_2$ are material coefficients. The usual form of the
Ginzburg-Landau free energy density is
\begin{equation}
    f(\xi, \nabla\xi)= f_0(\xi) - \gamma(\nabla\xi)^2/2,
\label{GL_fun}\end{equation}

\noindent where $\gamma$ is a material coefficient, $f_0$ is the static
(equilibrium) free energy and so $\Gamma_\xi = f'(\xi)$. Gurtin gave a
method to deduce equation (\ref{GL_eq}) from pure thermodynamic
considerations, with the concept of microforce balance and showed that
some additional terms should appear with a characteristic structure
\cite{Gur96a,Gur00b}. The generalized form of the equation together
with the characteristic term is the following
\begin{equation}
\partial_t\xi = -\gamma_1 \Gamma_\xi + \gamma_2 \Delta\xi +
\gamma_3\Delta\partial_t \xi,
\label{GL_eqgen}\end{equation}

Such kind of terms appear in connection of several different phenomena
and not only in case of the Ginzburg-Landau equation
\cite{Aif80a1,UngAta02a}. E. g. the Guyer-Krumhansl equation of heat
conduction can be considered as a Cattaneo-Vernotte type wave heat
conduction equation supplemented by a Gurtin term.

It was argued that the Ginzburg-Landau equation is the first nonlocal
extension of any kind of equation for an internal variable
\cite{Van02a1,Van03a1} and its characteristic functional form can be
derived from the requirement of compatibility with the Second Law,
without referring to variational principles. Hence, the reason of its
wide-range applicability is well founded, because any internal variable
that can characterize a material structure and is independent of other
requirements should fulfill a Ginzburg-Landau equation in the first
nonlocal approximation. The arguments were supported by calculations
based on Liu's theorem. However, in an internal variable, completely
relocalized theory one cannot derive directly (\ref{GL_eq}), but only a
very similar equation, that was called {\em thermodynamic
Ginzburg-Landau equation}
\begin{equation}
    \partial_t\xi = -\gamma_{T1} \Gamma_\xi
    + \gamma_{T2}  \Delta \Gamma_\xi,
\label{tGL_eq}\end{equation}

\noindent where $\gamma_{T1}$ and $\gamma_{T2}$ are material
coefficients. Moreover, the entropy (free energy) was proved to be
gradient independent. One can see, that the equations (\ref{GL_eq}) and
(\ref{tGL_eq}) are similar but not the same at all. The essential
qualitative difference is that static (equilibrium) solutions of the
thermodynamic Ginzburg-Landau equation with homogeneous boundary
conditions are homogeneous but static solutions of the original
Ginzburg-Landau equation with the same boundary conditions are not,
they can form structures. The situation is well known and understood in
superconductors. The London equation (corresponding to the
thermodynamic Ginzburg-Landau) does not determine the penetration
length of the magnetic field, however, the Ginzburg-Landau equation
gives that \cite{LanPit82b}.

In the following, applying Liu's procedure with the methodology
described in the previous section, we will derive the Ginzburg-Landau
equations from very general assumptions and show that the Gurtin terms
are consequences of pure thermodynamic considerations, without
referring new concepts, like the configurational force balance or
virtual power, etc..

We are looking for a dynamic equation of $\xi$ in the following general
form
\begin{equation}
 \partial_t\xi - \mathcal{F} = 0,
\label{GL_start}\end{equation}

\noindent where $\mathcal{F}$ is a constitutive function, which
form is to be restricted by the Second Law. The basic state space
is spanned by $\xi$. Let us assume that the constitutive space is
spanned by $\xi$, $\nabla \xi$ and $\nabla^2\xi$. In this case the
entropy inequality will be
\begin{eqnarray*}
\partial_ts &+& \nabla\cdot {\bf j}_s = \\
    \partial_1s \, \partial_t\xi &+&
    \partial_2s \cdot\nabla\partial_t \xi +
    \partial_3s :\nabla^2\partial_t \xi +
    \partial_1 {\bf j}_s \cdot \nabla{\xi} +
    \partial_2 {\bf j}_s : \nabla^2{\xi} +
    \partial_3 {\bf j}_s \cdot :\nabla^3{\xi} \geq 0.
\end{eqnarray*}

One can see, that the space of the process directions (independent
variables) is spanned by $\partial_t\xi$, $\nabla\partial_t\xi$,
$\nabla^2\partial_t\xi$ and $\nabla^3\xi$. Moreover, let us observe
that these variables are not really independent, the gradient of
(\ref{GL_start}) connect them. Therefore, in addition to
(\ref{GL_start})  one should consider the following constraint
\begin{equation}
\nabla\partial_t\xi + \nabla\mathcal{F} =
    \nabla\partial_t\xi +
    \partial_1\mathcal{F} \nabla{\xi} +
    \partial_2\mathcal{F}\cdot{\nabla^2{\xi}} +
    \partial_3 \mathcal{F}:\nabla^3{\xi} = 0.
\label{dGL_start}\end{equation}

Introducing $\Gamma_1$ and $\Gamma_2$ Lagrange-Farkas multipliers
for the constraints (\ref{GL_start}) and (\ref{dGL_start})
respectively, one can get the following Liu equations
\begin{eqnarray*}
\partial_1 s  &=& \Gamma_1, \\
\partial_2 s &=& \Gamma_2, \\
\partial_3 s &=& {\bf 0}, \\
(\partial_3 {\bf j}_s -\Gamma_2 \partial_3\mathcal{F})^s &=& {\bf 0}.
\end{eqnarray*}

Here the superscript $^s$ denotes the symmetric part of the
corresponding tensor. The first two equations determine the
multipliers. From the third equation follows, that the entropy does not
depend on the second derivative of $\xi$. Taking into account these
requirements one can give a solution of the fourth equation and
determine the entropy current as
\begin{equation}
{\bf j}_s(\xi,\nabla\xi,\nabla^2\xi) =
    \partial_2s(\xi,\nabla\xi)
        \mathcal{F}(\xi,\nabla\xi,\nabla^2\xi)
    + {\bf j}_0(\xi,\nabla\xi).
\label{GL_js}\end{equation}

For the sake of clarity we explicitly denoted the variables of the
corresponding functions. With the above solution of the Liu equations
the dissipation inequality can be simplified considerably
\begin{equation}
\nabla\cdot{\bf j}_0 + (\nabla\cdot\partial_2 s -
\partial_1s)\cdot\mathcal{F} \geq 0.
\label{GL_eprod}\end{equation}

Assuming, that ${\bf j}_0\equiv 0$ one can give the general solution of
the above inequality. That solution can be interpreted by the well
known traditional method of irreversible thermodynamics, choosing
appropriate forces and currents. Therefore, $\mathcal{F}$ is the
constitutive quantity to be determined (thermodynamic current) and it
should be proportional to the given one (force)
\begin{equation}
\partial_t\xi = \mathcal{F}= L(\nabla\cdot\partial_2 s -
\partial_1s)
\label{gGL_eq}\end{equation}

\noindent with a nonnegative state dependent constitutive function $L$.
$s(\xi, \nabla\xi)$ is a given entropy function (determined from static
measurements). (\ref{gGL_eq}) is the Ginzburg-Landau equation, and one
can get back the very traditional (\ref{GL_eq}) form using the specific
entropy functional with a form like (\ref{GL_fun}) and dealing with a
strictly linear theory in a thermodynamic sense, where $L$ is a
constant function. The choice of the right thermodynamic potential
(e.g. entropy or free energy) depends on the boundary conditions
\cite{PenFif90a}.

If we do not restrict the space of the independent variables by
(\ref{dGL_start}), by the derivative of the original constraint, then
after an easy calculations one can get the following form of the
dissipation inequality
\begin{equation}
\nabla\cdot{\bf j}_0(\xi,\nabla\xi)
    -\partial_1s(\xi) \mathcal{F}(\xi,\nabla\xi,\nabla^2\xi) \geq
    0.
\label{TGL_eprod}\end{equation}

The Liu equations require, that the entropy must not depend on the
gradients of the basic state. However, the dissipation inequality still
can be solved, if one considers two additional physical requirements,
prerequisites of relocalizability \cite{Van03a1}.
\begin{enumerate}
\item $\xi$ is a dynamic variable in a thermodynamic sense therefore
$\xi$ is zero in equilibrium,
\item there is no entropy flow connected to the dynamic variable if its
value is zero.
\end{enumerate}

It was argued that these requirements are week from a physical point of
view. With these assumptions one can specify ${\bf j}_0$ in the entropy
current with the Ny\'\i{}ri-form as ${\bf j}_0(\xi,\nabla\xi) = {\bf
A}(\xi,\nabla\xi) \xi$, or equivalently ${\bf j}_0(\xi,\nabla\xi) =
\hat{\bf A}(\xi,\nabla\xi) \partial_\xi s$ \cite{Nyi91a1}, according to
the mean value theorem. Here the {\em current multipliers} {\bf A} or
$\hat{\bf A}$  are constitutive functions to be determined. With the
second form of ${\bf j}_0$ the dissipation inequality (\ref{TGL_eprod})
can be solved and gives the thermodynamic Ginzburg-Landau equation
(\ref{tGL_eq}). Introducing an additional, new dynamic variable one can
recover the additional Gurtin-term in the thermodynamic Ginzburg-Landau
equation \cite{Van02a1}. Let us observe, that the key assumption
determining the form of the additional entropy current was that the
entropy current should not affect the equilibrium solutions. Because
the entropy is a local function in this case, the special internal
variables, the current multipliers in a sense relocalize the
nonlocalities.

We can apply similar reasoning in the previous non relocalizable case,
too. However, first we should extend the previous constitutive state
space considering a derivative one order higher then before. Therefore,
let be our constitutive space spanned by $(\xi, \nabla \xi, \nabla^2
\xi, \nabla^3 \xi)$. After a short calculation we can recover the
validity of (\ref{GL_eprod}) in these new variables. The only
difference is that the final constitutive quantities e.g. ${\bf j}_0$
and $L$ will depend on the larger constitutive state. Now we assume
that ${\bf j}_0$ has the following form
$$
{\bf j}_0(\xi,\nabla\xi,\nabla^2\xi) = \hat{\bf
B}(\xi,\nabla\xi,\nabla^2\xi) \mathcal{F}(\xi,\nabla\xi,\nabla^2\xi),
$$

\noindent were $\hat{\bf B}$ is a current multiplier. The above form is
a direct application of the requirement that the entropy current should
not change the equilibrium solutions (with some minor additional
restrictions on the possible constitutive dependencies). In this case
the entropy production (\ref{GL_eprod}) is
\begin{equation}
\hat{\bf B}\cdot \nabla\mathcal{F} + (\nabla\cdot\hat{\bf B} +
\nabla\cdot\partial_2 s -
\partial_1s)\cdot\mathcal{F} \geq 0.
\end{equation}

With two undetermined constitutive functions ($\hat{\bf B},
\mathcal{F}$) the inequality has a general solution, the currents and
forces are determined by the constitutive dependencies. In case of
isotropic materials
\begin{eqnarray*}
\hat{\bf B} &=& L_1\nabla\mathcal{F} \\
\mathcal{F} &=& L_2 (\nabla\cdot\hat{\bf B} +
\nabla\cdot\partial_2 s - \partial_1s).
\end{eqnarray*}

Here $L_1$ and $L_2$ are nonnegative scalar constitutive functions.
$\hat{\bf B}$ can be eliminated from the above equations and finally we
get
\begin{equation}
\partial_t\xi = \mathcal{F} = L_2(\nabla\cdot\partial_2 s -
\partial_1s) + L_2 \nabla\cdot(L_1 \nabla\partial_t \xi).
\label{egGL_eq}\end{equation}

This is the Ginzburg-Landau equation with a characteristic additional
Gurtin-term.

\section{Conclusions and discussion}

The requirement of a nonnegative entropy production is a relatively
strong and not a complete form of the Second Law. Strong form because
it is a local requirement and other weaker formulations require only
the validity of integral inequalities. Not a complete one because an
increasing entropy is only a part of the physical content of the Second
Law. The stability of materials in isolated systems incorporates some
other conditions (e.g. concave entropy function), too \cite{Mat00a}.
Coleman-Mizel methodology is a kind of basic philosophical requirement
of a thermodynamic theory: the acceptable theories are those, where the
entropy inequality is the consequence of pure material properties and
independent of other elements of the theory (e.g. initial conditions)
ensuring a kind of universality and some stability properties to any
thermodynamic theories.

It is interesting to know, that the doubled variational-thermodynamic
structure of the Ginzburg-Landau equation can be generalized
considerably. That is the idea behind the General Equation for the
Nonequilibrium Reversible-Irreversible Coupling (GENERIC)
\cite{GrmOtt97a,OttGrm97a,Ott98a}, where the variational part and a
formalism from mechanics plays the leading role (different brackets,
geometrical point of view, etc..), but both parts are represented. In
this paper we unified the variational and the thermodynamic parts of
the derivation of the Ginzburg-Landau equation on a pure thermodynamic
ground, where we did not refer to any kind of variational principle.
However, the derived static part turned out to have a complete
Euler-Lagrange form. The dynamic part contains a first order time
derivative therefore one cannot hope to derive it from a variational
principle of Hamiltonian type \cite{VanMus95a}. In our approach we get
the "reversible", "variational" part as a specific case of the
thermodynamic, irreversible thinking, but one cannot hope the contrary,
the irreversible part cannot be derived from a variational, reversible
thinking.

Weakly nonlocal, pattern forming equations emerge in different fields
of physics independently of thermodynamic argumentation. Understanding
their compatibility with the Second Law can be considered as one of the
most important challenges of contemporary non-equilibrium
thermodynamics. In this paper it was shown that one of the most
important pattern forming equations, the Ginzburg-Landau equation, is a
straightforward consequence of the entropy inequality alone in a
nonlocally extended constitutive space. On the other hand the outline
of the mathematical background and all the key ingredients of an
efficient formalism to exploit the Second Law inequality in weakly
nonlocal continuum theories (gradient theories, theories with coarse
grained thermodynamic potentials, phase-field models, etc..) is given.

\section{Appendix: Liu theorem as a variant of Farkas' lemma and some
of its consequences}

In 1972 Liu introduced a method of the exploitation of the entropy
principle \cite{Liu72a}. Liu's procedure became a basic tool to find
the restrictions posed by the entropy inequality. The method is based
on a linear algebraic theorem, called Liu's theorem in the
thermodynamic literature \cite{MulRug98b,MusAta01a} and on an
interpretation of the entropy inequality, one of the fundamental
ingredients of the Second Law. Recently Hauser and Kirchner recognized
that Liu's theorem is a consequence of the fundamental theorem of
linear inequalities, a famous statement of optimization theory and
linear programming, the so called Farkas' lemma \cite{HauKir02a}. That
theorem was proved first by Farkas in 1894 \cite{Far894a} and
independently by Minkowski in 1896 \cite{Min896b}. In this appendix we
formulate and generalize Liu's theorem in a way that is best adapted
for our purposes and shows the whole train of thought from Farkas'
lemma to Liu's theorem giving a simple proof to every statement in
question.

Farkas' lemma can be formulated in several different forms, that are
more or less equivalent \cite{Roc70b,Sch98b}. Here we start from a
simple variant.

\begin{lem}{(Farkas)}\label{Farlem}
Let ${\bf a}_i\neq {\bf 0}$ be independent vectors in a  finite
dimensional vector space $\mathbb{V}$, $i=1...n$, and $S = \{ {\bf p}
\in \mathbb{V}^* | {\bf p}\cdot{\bf a}_i \geq 0, i=1...n \}$. The
following statements are equivalent for a ${\bf b} \in \mathbb{V}$:

(i) $ {\bf p}\cdot {\bf b} \geq 0$, for all ${\bf p} \in S$.

(ii) There are nonnegative real numbers $\lambda_1, ..., \lambda_n$
such that ${\bf b} = \sum^n_{i=1} \lambda_i {\bf a}_i$.
\end{lem}

Proof: $S$ is not empty. In fact, for all $k, i \in \{1,...,n\}$ there
is a ${\bf p}_k\in \mathbb{V}^*$ such that ${\bf p}_k\cdot{\bf a}_k =
1$ and ${\bf p}_k\cdot{\bf a}_i = 0$ if $i\neq k$. Evidently ${\bf
p}_k\in S$ for all $k$.

$(ii) \Rightarrow (i)$ ${\bf p}\cdot \sum^n_{i=1} \lambda_i {\bf a}_i =
\sum^n_{i=1} \lambda_i {\bf p} \cdot {\bf a}_i \geq 0$ if ${\bf p} \in
S$.

$(i) \Rightarrow (ii)$ Let $S_0 = \{ {\bf y} \in \mathbb{V}^* | {\bf
y}\cdot{\bf a}_i = 0, i=1...n \}$. Clearly $\emptyset \neq S_0 \subset
S$.

If  ${\bf y}\in S_0$ then $-{\bf y}$ is also in $S_0$, therefore ${\bf
y}\cdot{\bf b}\geq 0$ and $-{\bf y}\cdot{\bf b}\geq 0$ together.
Therefore for all ${\bf y}\in S_0$ it is true that ${\bf y}\cdot{\bf
b}= 0$.

As a consequence ${\bf b}$ is in the set generated by $\{{\bf a}_i\}$,
that is  there are real numbers $\lambda_1,..., \lambda_n$ such that
${\bf b} = \sum^n_{i=1} \lambda_i{\bf a}_i$. These numbers are
nonnegative, because with the previously defined ${\bf p}_k\in S$,
$0\leq {\bf p}_k\cdot{\bf b} = {\bf p}_k \cdot \sum^l_{i=1}\lambda_i
{\bf a}_i = \lambda_i {\bf p}_k\cdot {\bf a}_i = \lambda_k$ is valid
for all $k$. $\blacksquare$

\begin{rem}
In the following the elements of $\mathbb{V}^*$ are called {\em
independent variables} and $\mathbb{V}^*$ itself is called the {\em
space of independent variables}. The inequality in the first statement
of the lemma is called {\em aim inequality} and the nonnegative numbers
in the second statement are called {\em Lagrange-Farkas multipliers}.
The inequalities determining $S$ are the {\em constraints}.

In the calculations an excellent reminder is to use Lagrange-Farkas
multipliers similarly to Lagrange multipliers in case of conditional
extremum problems:
$$
{\bf p}\cdot{\bf b} - \sum^n_{i=1}\lambda_i {\bf p}\cdot
    {\bf a}_i
= {\bf p}\cdot\left({\bf b} - \sum^n_{i=1}\lambda_i \cdot {\bf
    a}_i\right) \geq 0,
\quad \forall {\bf p}\in \mathbb{V}^*
$$

From this form we can read out the second statement of the lemma.
\end{rem}

\begin{rem}
The original statement does not require the independency of the vectors
in the constraint. We need some extra conditions and that
generalization destroys the simplicity of the proof. However, we do not
need this generalization in thermodynamics.

The geometric interpretation of the theorem is important and graphic:
either the vector ${\bf b}$ belongs to the cone generated finitely  by
the vectors ${\bf a}_i$ ($Cone({\bf a}_1, ..., {\bf a}_n) =
\{\lambda_1{\bf a}_1 + ... + \lambda_n {\bf a}_n \left| \right.
(\lambda_1,..., \lambda_n) \in \mathbb{R}^{+n}$), or there exists a
hyperplane separating ${\bf b}$ from the cone.
\end{rem}

\subsection{Affine Farkas' lemma}

This generalization of the previous lemma was first published
simultaneously by A. Haar and J. Farkas in the same number of the same
journal, with different proofs \cite{Haa18a,Far18a2}. Later it was
reproved independently by others several times (e.g.
\cite{Neu??m,Sch98b}). Here we give a simple version again.

\begin{thm}{(Affine Farkas)} Let ${\bf a}_i\neq {\bf
0}$ be independent vectors in a finite dimensional vector space
$\mathbb{V}$ and $\alpha_i$ real numbers, $i=1...n$ and  $S_A = \{ {\bf
p} \in \mathbb{V}^* | {\bf p}\cdot{\bf a}_i \geq \alpha_i, i=1...n \}$.
The following statements are equivalent for a ${\bf b}\in \mathbb{V}$
and a real number $\beta$:

(i) ${\bf p}\cdot {\bf b} \geq \beta$, for all ${\bf p} \in S_A$.

(ii) There are nonnegative real numbers $\lambda_1,...,\lambda_n$ such
that ${\bf b} = \sum^n_{i=1} \lambda_i {\bf a}_i$ and $\beta\leq
\sum^n_{i=1} \lambda_i \alpha_i$.
\end{thm}

Proof: $S_A$ is not empty. In fact, $\alpha_i {\bf p}_k\in S_A$ for all
$k$ (${\bf p}_k\cdot{\bf a}_k = 1$ and ${\bf p}_k\cdot{\bf a}_i = 0$ if
$i\neq k$ as previously).

$(ii) \Rightarrow (i)$ ${\bf p}\cdot{\bf b} = {\bf p}\cdot \sum^n_{i=1}
\lambda_i {\bf a}_i = \sum^n_{i=1} \lambda_i {\bf p} \cdot {\bf a}_i
\geq \sum^n_{i=1} \lambda_i \alpha_i \geq \beta$.

$(i) \Rightarrow (ii)$ First we will show indirectly that the first
condition of lemma  \ref{Farlem} is a consequence of the first
condition here, that is if (i) is true then ${\bf p}\cdot {\bf b} \geq
0$, for all ${\bf p} \in S$.

Thus let us assume the contrary, hence there is ${\bf p}'\in S$, for
which ${\bf p}'\cdot{\bf b} < 0$. Take an arbitrary ${\bf p}\in S_A$,
then ${\bf p} + k{\bf p}' \in S_A$ for all real numbers $k$. But now
$({\bf p} + k{\bf p}')\cdot {\bf b} = {\bf p}\cdot{\bf b} + k{\bf
p}'\cdot{\bf b} < \beta$, if $k \geq \frac{\beta - {\bf p}\cdot {\bf
b}}{{\bf p}'\cdot {\bf b}}$. That is a contradiction.

Therefore, according to Farkas' lemma (Lemma \ref{Farlem}) there exist
Lagrange-Farkas multipliers $\lambda = (\lambda_1,...,\lambda_n) \in
\mathbb{R}^{n+}$ such that ${\bf b} = \sum^n_{i=1} \lambda_i {\bf
a}_i$. Hence $\beta \leq inf_{p\in S_A} \{{\bf p}\cdot \sum^n_{i=1}
\lambda_i{\bf a}_i \} = inf_{p\in S_A} \{ \sum^n_{i=1} \lambda_i{\bf
p}\cdot {\bf a}_i \} = \sum^n_{i=1} \lambda_i \alpha_i$. $\blacksquare$

\begin{rem} The multiplier form is a good reminder again
$$
({\bf p}\cdot{\bf b} - \beta) - \sum^n_{i=1}\lambda_i ({\bf p}\cdot
{\bf a}_i -\alpha_i) = {\bf p}\cdot({\bf b} - \sum^n_{i=1}\lambda_i
\cdot {\bf a}_i) - \beta  + \sum^n_{i=1}\lambda_i \alpha_i \geq 0,
\quad \forall {\bf p}\in \mathbb{V}^*
$$.
\end{rem}

\begin{rem}
The geometric interpretation is similar to the previous one, but
everything is affine.
\end{rem}

\subsection{Liu's theorem}

Here the constraints are equalities instead of inequalities, therefore
the multipliers are not necessarily positive.

\begin{thm}{(Liu)}\label{Liu}
Let ${\bf a}_i\neq {\bf 0}$ be independent
vectors in a finite dimensional vector space $\mathbb{V}$ and
$\alpha_i$ real numbers, $i=1...n$ and $S_L = \{ {\bf p} \in
\mathbb{V}^* | {\bf p}\cdot{\bf a}_i = \alpha_i, i=1...n \}$. The
following statements are equivalent for a ${\bf b} \in \mathbb{V}$ and
a real number $\beta$:

(i) ${\bf p}\cdot {\bf b} \geq \beta$, for all ${\bf p} \in S_L$,

(ii) There are real numbers $\lambda_1,...,\lambda_n$ such that
\begin{equation}
    {\bf b} = \sum^n_{i=1} \lambda_i {\bf a}_i
\label{Liu_eq}\end{equation} and
\begin{equation}
    \beta\leq \sum^n_{i=1} \lambda_i \alpha_i.
\label{dis_ineq}\end{equation}
\end{thm}

Proof: A straightforward consequence of the previous affine form of
Farkas' lemma because $S_L$ can be given in a form $S_A$ with the
vectors ${\bf a}_i$ and $-{\bf a}_i$, $i=1,...,n$: $S_L = \{ {\bf p}
\in \mathbb{V}^* | {\bf p}\cdot{\bf a}_i \geq \alpha_i, \quad
\text{and}\quad {\bf p}\cdot(-{\bf a}_i) \geq -\alpha_i, i=1...n \}$.

Therefore there are nonnegative real numbers
$\lambda_1^+,...,\lambda_n^+$ and $\lambda_1^-,...,\lambda_n^-$ such,
that ${\bf b} = \sum^n_{i=1} (\lambda^+_i {\bf a}_i - \lambda^-_i {\bf
a}_i) = \sum^n_{i=1} (\lambda^+_i - \lambda^-_i) {\bf a}_i =
\sum^n_{i=1} \lambda_i {\bf a}_i $ and $\beta \leq \sum^n_{i=1}
(\lambda^+_i \alpha_i - \lambda^-_i \alpha_i)$. $\blacksquare$

\begin{rem} The multiplier form is a help in the applications
again
$$
0\leq  ({\bf p}\cdot{\bf b} - \beta) - \sum^n_{i=1} \lambda_i
    ({\bf p}\cdot {\bf a}_i -\alpha_i)
= {\bf p}\cdot({\bf b} - \sum^n_{i=1}\lambda_i \cdot {\bf a}_i)
    -\beta  + \sum^n_{i=1} \lambda_i \alpha_i,
\quad \forall {\bf p}\in \mathbb{V}^*.
$$
\end{rem}

\begin{rem} In the theorem with Lagrange multipliers for a local
conditional extremum of a differentiable function we apply exactly the
above theorem of linear algebra after a linearization of the
corresponding functions at the extremum point.
\end{rem}

Considering the requirements of the applications we generalize Liu's
theorem to take into account vectorial constraints. First of all let us
remember some well known identifications of linear algebra:
$Lin(\mathbb{U}^*,\mathbb{V}) \equiv Bilin(\mathbb{U}^*
\times\mathbb{V}^*,\mathbb{R}) \equiv \mathbb{V}\otimes\mathbb{U}$,
where $Bilin$ denotes the bilinear mappings of the corresponding spaces
(see e.g. \cite{Mat93b}).

\begin{thm}{(vector Liu)}\label{vectLiu}
Let ${\bf A}\neq {\bf 0}$ in a tensor product $\mathbb{V}\otimes
\mathbb{U}$ of finite dimensional vector spaces $\mathbb{V}$ and
$\mathbb{U}$. Let $\boldsymbol{\alpha}$ $\in \mathbb{U}$ and $S_{VL} =
\{ {\bf p} \in \mathbb{V}^* | {\bf p}\cdot{\bf A} = {\bf \alpha}\}$.
The following statements are equivalent for a ${\bf b} \in \mathbb{V}$
and a real number $\beta$:

(i) ${\bf p}\cdot {\bf b} \geq \beta$, for all ${\bf p} \in S_{VL}$.

(ii) There is a $\boldsymbol{ \lambda}\in \mathbb{U}^*$ such that
\begin{equation}
    {\bf b} = {\bf A}\cdot \boldsymbol{\lambda},
\label{Liu_veq}\end{equation} and
\begin{equation}
    \beta\leq \boldsymbol{ \lambda}\cdot \boldsymbol{\alpha}.
\label{dis_vineq}\end{equation}
\end{thm}

Proof: Let us observe that we can get back the previous form of the
theorem by introducing a linear bijection ${\bf K}: \mathbb{U}
\rightarrow \mathbb{R}^n$, a {\em coordinatization} in $\mathbb{U}$.
Therefore, applying it for ${\bf K}\cdot {\bf A} = ({\bf A})_i = {\bf
a}_i$, ${\bf K}\cdot \boldsymbol{\alpha} = (\boldsymbol{\alpha})_i =
\alpha_i$ and ${\bf K}'\cdot {\bf A} = {\bf a}'_i$, ${\bf K}'\cdot
\boldsymbol{\alpha} = \alpha'_i$ we get that ${\bf b} = \sum^n_{i=1}
\lambda_i {\bf a}_i = \sum^n_{i=1} \lambda'_i {\bf a}'_i$. Thus
$\lambda'_i = {\bf K}'^{*-1}\cdot{\bf K}^*\cdot \lambda_i$. Therefore
there is a $\boldsymbol{\lambda}\in \mathbb{U}$, independently of the
coordinatization, with the components $\lambda_i$ and $\lambda'_i$ in
the coordinatizations ${\bf K}$ and ${\bf K}'$. $\blacksquare$

The previously excluded degenerate case of ${\bf A}={\bf 0}$ deserves a
special attention. Now we require the validity of the aim inequality
for all ${\bf p}\in \mathbb{V}^*$ without any constraint. The
consequences can be formulated as previously and the proof is trivial.

\begin{thm}{(degenerate Liu)}\label{degLiu}
The following statements are equivalent for a ${\bf b} \in \mathbb{V}$
and a real number $\beta$:

(i) ${\bf p}\cdot {\bf b} \geq \beta$ for all ${\bf p}\in
\mathbb{V}^*$.

(ii)  ${\bf b} = {\bf 0}$ and $\beta\leq 0$.
\end{thm}

\begin{rem}
The practical application rule is that if ${\bf A}={\bf 0}$ then the
multiplier is zero.
\end{rem}

\begin{rem} In continuum physics and thermodynamics the
corresponding form of (\ref{Liu_veq}) and (\ref{dis_vineq}) are called
{\em Liu equation(s)} and the {\em dissipation inequality},
respectively. We apply the same names for the degenerate case, too.
There the Lagrange-Farkas multipliers are called simply Lagrange
multipliers. Our nomenclature honors Farkas and emphasizes the
difference between the two kind of multipliers. It can be important
also to make a clear distinction of a similar but different
nomenclature and method in variational principle construction in
continuum physics \cite{SieBer02a}.
\end{rem}

\section{Acknowledgements}

Thanks for Raphael Hauser, Vincent Koroneos, Tam\'as Matolcsi and
Christina Papenfuss, for valuable critical remarks and careful reading
of the manuscript and for Antonio Cimmelli for interesting discussions.
This research was supported by OTKA T034715 and T034603.

\end{document}